*Title:* A quantile-based g-computation approach to addressing the effects of exposure mixtures

*Authors*:
Alexander P. Keil[a,b], Jessie P. Buckley[c], Katie M. O'Brien[b], Kelly K. Ferguson[b], Shanshan Zhao[d]
Alexandra J. White [b]

[a] Department of Epidemiology, University of North Carolina, Chapel Hill, North Carolina
[b] Epidemiology Branch, National Institute of Environmental Health Sciences (NIEHS), National Institutes of Health (NIH), Department of Health and Human Services (DHHS), Research Triangle Park, North Carolina, USA.
[c] Department of Environmental Health and Engineering, and Department of Epidemiology, The Johns Hopkins University, Baltimore, Maryland
[d] Biostatistics Branch, National Institute of Environmental Health Sciences (NIEHS), National Institutes of Health (NIH), Department of Health and Human Services (DHHS), Research Triangle Park, North Carolina, USA.

*Corresponding Author:*
Alexander Keil, Department of Epidemiology, CB 7435, University of North Carolina, Chapel Hill, NC 27599-7435. E-mail: akeil@unc.edu. T: 919-966-6652. F: 919-966-2089

The authors report no conflicts of interest.

This research was supported in part by the NIH/NIEHS grant #s R01ES029531, R01ES030078 and the Intramural Research Program of the NIH, NIEHS #Z01ES044005

ABSTRACT

Exposure mixtures frequently occur in data across many domains, particularly in the fields of environmental and nutritional epidemiology. Various strategies have arisen to answer questions about exposure mixtures, including methods such as weighted quantile sum (WQS) regression that estimate a joint effect of the mixture components.

We demonstrate a new approach to estimating the joint effects of a mixture: quantile g-computation. This approach combines the inferential simplicity of WQS regression with the flexibility of g-computation, a method of causal effect estimation. We use simulations to examine whether quantile g-computation and WQS regression can accurately and precisely estimate effects of mixtures in a variety of common scenarios.

We examine the bias, confidence interval coverage, and bias-variance tradeoff of quantile g-computation and WQS regression, and how these quantities are impacted by the presence of non-causal exposures, exposure correlation, unmeasured confounding, and non-linearity of exposure effects.

Quantile g-computation, unlike WQS regression allows inference on mixture effects that is unbiased with appropriate confidence interval coverage at sample sizes typically encountered in epidemiologic studies, and when the assumptions of WQS regression are not met. Further, WQS regression can magnify bias from unmeasured confounding that might occur if important components of the mixture are omitted from analysis. Unlike inferential approaches that examine effects of individual exposures, while holding other exposures constant, methods like quantile g-computation that can estimate the effect of a mixture are essential for understanding effects of potential public health actions that act on exposure sources. Our approach may serve to help bridge gaps between epidemiologic analysis and interventions such as regulations on industrial emissions or mining processes, dietary changes, or consumer behavioral changes that act on multiple exposures simultaneously.

INTRODUCTION

Epidemiologists are increasingly confronted with arrays of unique, yet sometimes highly correlated, exposures of interest that may arise from similar sources and provide unique challenges to inference. The myriad analytic methods for mixtures data have been subject to a number of recent reviews and workshops about mixtures data (1-3). A clear theme has emerged that numerous questions can be answered in mixtures data, and different methods are best suited for different questions. For example, we might be interested in questions about the mixture (clustering of exposures, projecting exposures to a lower dimensional space) or questions about the health effects of individual components of the mixture, or of joint effects of multiple (perhaps all) components of the mixture. Here, we focus in particular on one of these many questions that can be asked of epidemiologic data on exposure mixtures: how can the mixture as a whole, rather than individual components, influence health of the populations exposed to the multitude of components in the mixture? Each of these components may act independently, synergistically, antagonistically, or they may be inert with respect to the health outcomes of interest. Focusing on the mixture as a whole is advantageous as it can provide simplicity of inference, integrate over multiple exposures that likely originate from similar sources, and often map directly onto the effects of potential public health interventions (4). The "mixture effect" is a useful metric even in light of policies that target single exposures. Particulate matter, for example, is a target of regulation by the US Environmental Protection Agency, but the actual interventions that have reduced particulate matter occur on the sources of pollution, and such interventions rarely act on only one element of the mixture of air pollutants. Thus, a better quantification of the effects of reducing particulate matter would account for joint reductions of other pollutants from particulate matter sources, such as sulfur dioxide and nitrogen oxides (from sources such as coal-fired power plants) or ozone and carbon monoxide (from sources such as automobile emissions).

One of the analytic approaches developed specifically for estimating effects of exposure mixtures, weighted quantile sum (WQS) regression, has become increasingly used as an analytic approach for exposure mixtures in relation to health outcomes (5-7). This approach is based on developing an exposure index that is a weighted average of all exposures of interest, after each exposure is transformed into a categorical variable defined by quantiles of the exposures. The index, representing the exposure mixture as a whole, is then used in a generalized linear model to estimate associations with health outcomes. WQS regression has a specific goal of estimating the effect of the mixture as a whole. While several methods from the causal inference literature are capable of estimating the effect of a whole mixture (4, 8-10), WQS regression has the advantage over these methods of a simple implementation. Although such approaches from the causal inference literature (e.g. inverse probability weighting and g-computation) can be implemented with standard regression software (11-13), WQS regression is available in a single, self-contained R function that only requires specification of a single model (14).

Despite the attractive features of this approach, there are two notable reasons to be cautious when applying WQS regression. First, WQS regression requires what we call a "directional homogeneity" assumption, which assumes that all exposures have co-adjusted associations with the outcome that are in the same direction (or can be coded, *a priori*, to meet this assumption) or are null. Second, WQS regression also assumes the individual exposures have linear and additive effects. Little is known about the demonstrable benefits of these assumptions when they are true and whether there are adverse impacts of these assumptions in realistic epidemiologic data where such assumptions would never be met exactly. Third, little theoretical statistical framework (15) and few simulations (14) exist that assess the internal validity of effect estimation (e.g. bias, confidence interval coverage) of WQS. While other statistical approaches have been developed to estimate overall mixture effects that are discussed in previously noted reviews and workshops, we know of no methods other than WQS regression that can provide parsimonious parametric inference for the effect of a mixture of exposures (e.g. a low-dimensional mixture dose-response such as a single coefficient corresponding to the change in the outcome per unit of joint exposure). Thus, there is need for additional methods that give similar inference without making such strong assumptions, as well as an analysis of the conditions under which such assumptions are warranted and possibly beneficial.

In the current manuscript, we demonstrate a new approach to estimating the effects of an exposure mixture, which we call quantile g-computation, that shares the simplicity of interpretation and computational ease of WQS regression while not assuming directional homogeneity. Further, our approach inherits many features of causal inference methods which allow for non-linearity and non-additivity of the effects of individual exposures and the mixture as a whole. We note explicit connections between these two approaches and demonstrate when they give equivalent estimates. We compare, using simulations, the validity of our approach and WQS regression for hypothesis testing, estimating the effects of a mixture, the control of confounding by correlated exposures, the impact of unmeasured confounding, and the impact of non-linearity and non-additivity on effect estimates.

METHODS

We first describe WQS regression in relation to standard generalized linear models. We then introduce quantile g-computation as a generalization and extension to WQS regression.

*WQS regression*
WQS regression developed gradually out of methods designed to estimate the relative contributions of exposures in a mixture to a single health outcome. WQS regression requires inputs similar to a standard regression model: an outcome (Y), a set of exposures of interest ($\boldsymbol{X}$) and a set of other covariates or confounders of interest ($\boldsymbol{Z}$). WQS starts by transforming $\boldsymbol{X}$ into a set of categorical variables in which the categories are created using quantiles of $\boldsymbol{X}$ (i.e. quantiles of each column) as cut points, which we denote as $\boldsymbol{X^q}$. The output of WQS regression consists of two parts: a regression model between the outcome of interest and an "index" exposure (possibly adjusting for covariates), and a set of weights that describe the contribution of each individual exposure to the single "index" exposure and overall effect estimate.

For continuous Y measured in individual $i$, the regression model part of WQS regression can be expressed as
$$Y_i = \beta_0 + \psi S_i + \epsilon_i$$

Where $\beta_0$ is the model intercept, $S_i$ is the exposure index (defined below), $\psi$ is the coefficient representing the incremental change in the expected value of $Y$ per unit increase in the $S_i$ and $\epsilon_i$ is the error term. The exposure index is defined as in **Equation 1** where $w_j$ are the "weights" for each exposure (here we have $d$ exposures) and $X_{ji}^q$ is the "quantized" version of the $j$th exposure for the $i$th individual.

$$(1) \quad S_i = \sum_{j=1}^{d} w_j X_{ji}^q$$

That is, if we use quartiles, then $X_{ji}^q$ will equal 0, 1, 2, or 3 for any participant, corresponding to whether the exposure $X_{ji}$ falls into the 0-25th, 25th-50th, 50th-75th, or 75th-100th percentile of that exposure. The weights are estimated as the mean weight across bootstrap samples of the WQS regression model (often in a distinct training set, which is discussed below) with all weights forced to sum to 1.0 and have the same sign (referred to as a "non-negativity/non-positivity constraint", which is enforced via constrained optimization or post-hoc selection of positive coefficients from an unconstrained model (15)). These constraints imply that all exposures contribute to $S_i$ in the same direction. As a consequence, in order for $\psi$ to be an unbiased and consistent estimator of an overall effect of $X^q$ on $Y$, we must assume directional homogeneity: all exposures must have the same direction of effect (inclusive of the null) with the outcome.

For intuition purposes, we can express a WQS regression as a standard linear model for the quantized exposures, given by **Equation 2**

$$(2) \quad Y = \beta_0 + \psi \sum_{j=1}^{d} w_j X_j^q + \epsilon = \sum_{j=1}^{d} \beta_j X_j^q + \epsilon$$

where we have suppressed subscript notation for individuals for clarity. In fact, if the directional homogeneity assumption holds, then all $\beta_j$ are, in fact positive and need not be constrained (ignoring sampling variability) and the WQS regression approach is equivalent to a generalized linear model in very large samples (as suggested by (15)). This equivalence is useful because generalized linear models do not require directional homogeneity, so this suggests, at the very least, that WQS regression will have similar inference to a generalized linear model, so generalized linear models might form the basis of an alternative to WQS regression.

There is little guidance on exactly how WQS regression results can be interpreted because "index effect" is not well-defined in the sense that it does not map onto real-world quantities. However, we can make some progress on what WQS estimates in large samples when the underlying model (**Equation 2**) is correct and exposure effects are in the same direction. We start by noting that an increase in the exposure index value from $S_i$ to $S_i + 1$ (the effect of which is estimated in a regression of the outcome on the exposure index) can be equated to a generalized linear model under directional homogeneity in the following derivation.

$$\psi = \psi(1 + S_i) - \psi S_i$$

$$= \psi \left( \sum_{j=1}^{d} w_j + \sum_{j=1}^{d} w_j X_{ji} \right) - \psi \sum_{j=1}^{d} w_j X_{ji}$$

$$= \left( \sum_{j=1}^{d} \psi w_j + \sum_{j=1}^{d} \psi w_j X_{ji} - \sum_{j=1}^{d} \psi w_j X_{ji} \right)$$

$$\text{(3)} \qquad = \sum_{j=1}^{d} \beta_j$$

Thus, by **Equation 3**, the mixture effect $\psi$ is equivalent to the sum of the $\beta$ coefficients in the underlying linear model, and the extension to alternative link functions is trivial. Such an interpretation suggests that we might be able to estimate $\psi$ using existing tools from causal inference. In small or moderate samples WQS regression may not act as the large sample results suggest, even under directional homogeneity. This occurs due to sampling variability and because WQS regression utilizes sample splitting, whereby it first estimates the weights in the training set and then estimates the mixture effect in the validation set, given those weights. Thus, small/moderate sample performance of WQS is poorly understood because it has not been explored sufficiently in the literature.

*Quantile g-computation*
One side effect of the large sample equivalence between generalized linear models and WQS regression (assuming directional homogeneity) is that, assuming no unmeasured confounding, WQS regression can be used to estimate causal effects if the linear model above is correct (which implies that quantization of $X$ is appropriate). This holds because generalized linear models are often used as the basis of causal effect estimation (e.g. (12, 16-23)) or the model parameters may be considered causal estimands (24). From an intervention perspective (and assumptions noted below), each $\beta_j$ is interpretable as the effect of increasing $X_j$ by one unit, so $\psi$ is interpretable as the effect of increasing all $X_j$ by one unit at the same time. In fact, $\psi$ is equivalent to the g-computation estimator (12) of a joint marginal structural model for quantized exposures, which estimates the effect of increasing every exposure simultaneously by one quantile. That is, $\psi$ can estimate a causal dose-response parameter of the entire exposure mixture. While g-computation has been described in detail by multiple authors in both frequentist and Bayesian settings (8, 11, 12, 25-27), we note here that it is a generalization of standardization that uses the law of total probability to compute estimates of the (usually population average) expected outcome distribution under specific exposure patterns. It is often combined with causal inference assumptions to select specific variables and models that, in practice, are used to predict, or simulate, outcome distributions that we would expect under different interventions on exposure that may depend on time-fixed or time-varying factors.

Under standard causal identification assumptions (including causal consistency, no interference and no unmeasured confounding, outlined in detail in (28)) and correct model specification, g-computation (or the g-formula), can yield the expected outcome, had we been able to intervene on all exposures of interest (4, 16). Under these assumptions, g-computation can be used to estimate causal effects of time-fixed or time-varying exposures. As described later in this section, g-computation can be used to estimate the parameters of a marginal structural model, which can quantify causal dose response parameters such as the change in the expected outcome expected as exposures are increased (i.e. by manipulating exposures as in an intervention). The causal assumptions are universal to all inferential approaches (29): any analysis should strive to measure all confounders and specify models as accurately as is reasonably possible. We mention them here in order be clear about how $\psi$ can be interpreted. In the special case of time-fixed exposures that enter into the model only with linear terms (additivity and linearity) the g-computation estimator of $\psi$ is given as the sum of all regression coefficients of the exposures of interest. $\psi$ corresponds to the change in $Y$ expected for a one-unit change in all exposures. Variance can be obtained using standard rules for estimating the sum of random variables and the covariance matrix of a linear model, which means that (in this simple setting) this approach requires little more computational time than a standard linear model.

When g-computation is performed using quantized exposures, we refer to the approach as "quantile g-computation." Quantile g-computation allows us to estimate both $\psi$ and the weights when the directional homogeneity assumption holds, but we will demonstrate that it allows valid inference regarding the effect of the whole exposure mixture, and individual contributions to that mixture, when directional homogeneity does not hold.

The first step of quantile g-computation is to transform the exposures $X_j$ into the quantized versions $X_j^q$. Second, we fit a linear model (where we have omitted other confounders $Z$ for notational simplicity, but they could also be included):

$$Y_i = \beta_0 + \sum_{j=1}^{d} \beta_j X_{ji}^q + \epsilon_i$$

Third, assuming directional homogeneity (for now), where $\psi$ is given as $\sum_{j=1}^{d} \beta_j$ (where $\beta_j$ is the effect size for exposure $j$) and the weights for each exposure (indexed by $k$) are given as $w_k = \beta_k / \sum_j^d \beta_j$ (i.e. the weights are defined to sum to 1.0). When directional homogeneity does not hold, quantile g-computation redefines the weights to be negative or positive weights, which are interpreted as the proportion of the negative or positive partial effect due to a specific exposure (and positive and negative weights are defined to sum to 1.0). When exposures may have non-linear effects on the outcome, we can extend this approach to include (for example) polynomial terms for $X_j^q$ in a model such as $Y_i = \beta_0 + \sum_{j=1}^{d} \beta_j X_{ji}^q + \sum_{j=1}^{d} \beta_{j+d} X_{ji}^q X_{ji}^q + \epsilon_i$ or a model that uses indicator variables for each level of a quantized exposure variable. Under non-linearity or when we include product terms between exposures (or between exposures and confounders), the weights themselves are not well defined (because the proportional contribution of an exposure to the overall effect would then vary according to levels of other variables and no longer function as weights) and $\psi$ is no longer simply a sum of the $\beta$ coefficients. However, $\psi$ is still easily estimable via standard g-computation algorithms ($\psi$ is a parameter of a marginal structural model estimated via g-computation, which generalizes the approach of Snowden et al. to a multilevel, joint exposure (12)), and the variance can be estimated with a non-parametric bootstrap. Briefly (as described in (12)), this approach involves the following steps: 1) fit an underlying model that allows individual effects of exposures on the outcome including interaction and non-linear terms; 2) make predictions from that model at set levels of the exposures; and 3) fit a marginal structural model to those predictions. Under linearity and additivity, this algorithm yields equivalent estimates of $\psi$ to the algorithm given above, but at a higher computational burden. This further extension allows the effect of the whole mixture be non-linear, as well (which is generally good practice if individual exposures are expected to have non-linear/non-additive effects). Extension to binary outcomes is based on a logistic model for Pr $(Y = 1|X^q)$, where $\psi$ can represent log-odds-ratios or log-risk-ratios, and this approach can also be extended to survival outcomes by using the Cox proportional hazards model as the underlying model to estimate hazard-ratios that quantify a mixture effect. These methods are implemented in the R package "qgcomp."

As we show below, quantile g-computation can be used to consistently estimate effects of the exposure mixture in settings in which WQS regression may to be biased or inconsistent but also yield equivalent estimates with WQS regression in large samples when its assumptions hold. In addition to avoiding the directional homogeneity assumption, quantile g-computation can be extended to settings in which the effects of exposures may not be additive (e.g. we might wish to include interaction terms among $X^q$) or non-linear (e.g. we may wish to include polynomial terms among $X^q$) and the exposure mixture effect may also be non-linear. WQS regression, on the other hand, assumes additivity among exposures and allows non-linear effects that are restricted to polynomial terms for $S_i$. The index exposure is still derived under a linear model, however, so it is not strictly equal to the quantile g-computation estimate under non-linearity of the mixture effect, which is based off of the fit of a marginal structural model and retains its interpretation as the effect of the mixture.

To demonstrate aspects of quantile g-computation and to compare the performance of quantile g-computation and WQS regression across a range of scenarios, we performed a number of simulation analyses that assess bias and confidence interval coverage for effect estimates, and power and type-I error for hypothesis tests.

*Simulation methods*
We simulate data on a mixture of $d$ exposures and a single, continuous outcome, where we let $d$ equal 4, 9, or 14, in sample sizes of 100 or 500 (and up to 5000, when noted). We simulate exposures such that they exactly equal the quantiles. For example, $X_1$ in each setting is simulated as a multinomial variable that takes on values 0, 1, 2, 3, each with probability 0.25, such that $X_1 = X_1^q$ in each case (q=4 in all scenarios). Note that this simulation scheme does not allow us to quantify potential bias from quantizing exposures that are measured on a continuous scale but enables us to isolate the factors we are interested in and avoid conflating bias from the estimation approach with bias from model misspecification.

Unless otherwise specified, we simulate the outcome according to the following model:
$$Y = \beta_0 + \beta_1 X_1 + \beta_2 X_2 + \epsilon = \beta_0 + \psi w_1 X_1 + \psi w_2 X_2 + \epsilon$$
All other exposures $X_3, \ldots, X_d$ are assumed to be "noise exposures" (except in scenario 4, below) with no effect on the outcome (equivalently: $\beta_3 = \ldots = \beta_d = 0$) and $\epsilon$ is simulated from a standard normal distribution (mean=0, standard deviation=1) for every scenario. The term "noise exposures" is used because, unless otherwise specified, they are not correlated with other exposures and do not contribute to the outcome. By varying d, we emulate exploratory studies of mixtures when some exposures may not affect the outcome, but the number of non-causal exposures may vary by context.

*Large sample simulation:* We demonstrate, empirically, that WQS regression (under directional homogeneity) is equivalent to a generalized linear model fit with quantized exposures (and thus quantile g-computation) in large samples. To demonstrate appropriate interpretation of WQS regression output and large sample equivalence between WQS regression and generalized linear when the directional homogeneity assumption, linearity, and additivity all hold, we performed a simulation in a single large dataset (N=100,000) with four exposures, $\psi = 5.0$ and $w_1 = \frac{\beta_1}{\psi} = 0.5, \frac{\beta_2}{\psi} = 0.25, \frac{\beta_3}{\psi} = 0.15, \frac{\beta_4}{\psi} = 0.1$. We analyze these data using WQS regression and quantile g-computation and report estimates of $\psi$, $\beta$, and $w$ for each approach.

*Small and moderate sample simulations*: We also contrast our new method with WQS regression in simulation settings with small or moderate sample sizes (i.e. typical sizes of observational studies) and when the necessary assumptions of WQS regression may be violated.

To address our study questions, we analyze data simulated under the scenarios given in **Table 1**, and described here:
1. $\beta_1 = \beta_2 = \psi = 0$. i.e. are null hypothesis tests ($H_0: \psi = 0$) valid in each approach when no exposures have effects on the outcome?
2. $\beta_1 = 0.25, \beta_2 = -0.25, \psi = 0$, i.e. are null hypothesis tests ($H_0: \psi = 0$) valid in each approach when exposures have counteracting effects on the outcome (directional homogeneity does not exist)?
3. $\beta_1 = 0.25, \beta_2 = 0, \psi = 0.25, w_1 = 1.0, w_2 = 0.0$, i.e. are estimates and confidence intervals valid in each approach when directional homogeneity does exist, and only a single exposure is causal?
4. $\beta_1 = \beta_2 = \cdots = \beta_d = \frac{0.25}{d}, \psi = 0.25, w_1 = w_2 = \cdots = w_p = \frac{1}{p}$, i.e. are estimates and confidence intervals valid in each approach when directional homogeneity does exist, and all exposures have causal effects?
5. $\beta_1 = 0.25, \beta_2 \in -0.2, -0.1, -0.05, \psi \in 0.05, 0.15, 0.25, \rho_{X_1,X_2} \in 0, 0.4, 0.75$, where $\rho_{X_1,X_2}$ is the Pearson correlation coefficient between $X_1$ and $X_2$. i.e. are estimates and confidence intervals valid in each approach when directional homogeneity does not exist due to negative co-pollutant confounding?
6. $\beta_1 \in 0, 0.25, \beta_2 = 0.0, \beta_C = 0.5, \psi \in 0, 0.25, \rho_{X,C} = 0.75$, where $\beta_C$ is the effect size of an unmeasured confounder $C$ and $\rho_{X,C}$ is the Pearson correlation coefficient between $X_1$ and $C$ (where $C$ is generated similarly to other exposures). I.e. what are the impacts of unmeasured confounding on the validity of each approach?

7. $\beta_1 = 0.25, \beta_2 = 0.25, \beta_{1,2} = -0.15, \psi_1 = 0.50, \psi_2 = -0.15$, where $\beta_{1,2}$ is the coefficient for the product term $X_1X_2$ in the underlying model, and $\psi_1$ and $\psi_2$ are the overall exposure coefficients for a linear and squared term for overall exposure. i.e. are estimates and confidence intervals valid in each approach when exposures interact on the model scale and the overall exposure effect is non-linear? For this analysis, the underlying model for quantile g-computation included a product term for $X_1X_2$ and the overall exposure effect was allowed to be non-linear by including linear and quadratic terms.
8. $\beta_1 = 0.25, \beta_2 = 0.25, \beta_{1,1} = -0.15, \psi_1 = 0.50, \psi_2 = -0.15$, where $\beta_{1,1}$ is the coefficient for the product term $X_1^2 = X_1X_1$ in the underlying model, and $\psi_1$ and $\psi_2$ are the overall exposure coefficients for a linear and squared term for overall exposure. i.e. are estimates and confidence intervals valid in each approach when exposures and the overall exposure effect are non-linear? For this analysis, the underlying model for quantile g-computation included a term for $X_1X_1$ and the overall exposure effect was allowed to be non-linear by including linear and quadratic terms.

*Simulating correlated, quantized exposures:* For scenario 6, we used a novel approach to induce a correlation ($\rho$) between two exposures $X_1$ and $X_2$ of $\rho_{X_1,X_2}$ by drawing values of $X_{2i} = X_{1i}$ with probability $\rho_{X_1,X_2}^{1/2}$, and otherwise drawing from other values of $X_1$. This sampling scheme ensured that exposures took on values of 0, 1, 2, or 3 with equal probability so that modeling assumptions would be met in all simulations. The simulation scheme was identical when considering unmeasured confounding, with the exception that we did not adjust for the unmeasured confounder in analyses.

**Table 1: Summary of simulation scenarios used to explore performance of quantile g-computation and WQS regression for small (N=100) or moderate (N=500) sized samples**

| Simulation scenario[a] | $\beta_1$ | $\beta_2$ | $\beta_{1,1}$ | $\beta_{1,2}$ | $\beta_C$ | $\psi_1$ | $\psi_2$ | $\rho_{X_1,X_2}$ | $\rho_{X,C}$ |
|---|---|---|---|---|---|---|---|---|---|
| 1 | 0 | 0 | 0 | 0 | 0 | 0 | 0 | 0 | 0 |
| 2 | 0.25 | -0.25 | 0 | 0 | 0 | 0 | 0 | 0 | 0 |
| 3 | 0.25 | 0 | 0 | 0 | 0 | 0.25 | 0 | 0 | 0 |
| 4[b] | 0.25/d | 0.25/d | 0 | 0 | 0 | 0.25 | 0 | 0 | 0 |
| 5 | 0.25 | -0.2, -0.1, -0.05 | 0 | 0 | 0 | 0.05, 0.15, 0.2 | 0 | 0.0, 0.4, 0.75 | 0 |
| 6 | 0.25 | 0 | 0 | 0 | 0.5 | 0.25 | 0 | 0 | 0.75 |
| 7 | 0.25 | 0.25 | 0 | -0.15 | 0 | 0.5 | -0.15 | 0 | 0 |
| 8 | 0.25 | 0.25 | -0.15 | 0 | 0 | 0.5 | -0.15 | 0 | 0 |

Table columns are as follows: $\beta_1$ = true coefficient for $X_1$, $\beta_2$ = true coefficient for $X_2$, $\beta_{1,2}$ = true coefficient for interaction term $X_1X_2$, $\beta_C$ = true coefficient for unmeasured confounder C, $\psi_1$= true mixture effect (main term), $\psi_2$ = true mixture effect (quadratic term), $\rho_{X_1,X_2}$ = true correlation between $X_1$ and $X_2$, $\rho_{X,C}$ = true correlation between $X_1$ and unmeasured confounder $C$

[a]Each scenario was repeated for sample sizes of 100 and 500 and a total number of exposures of 4, 9, and 14. Outcomes are simulated according to the model $Y = 0 + \sum_{j=1}^{d}\beta_jX_j + \beta_{1,1}X_1X_1 + \beta_{1,2}X_1X_2 + \beta_CC + \epsilon; \epsilon \sim N(0,1)$

[b]d refers to the total number of exposures

For all scenarios we simulated 1000 datasets and analyzed each of them using WQS regression (using the 'gWQS' (30) package defaults in R (31), weights assumed positive) and quantile g-computation (using the 'qgcomp' (32) package defaults in R). We report statistics relating to the $\psi$ parameter from each approach: bias (the mean estimate minus the true value of $\psi$, which is known in advance), the square root of the mean variance estimate from each method across the 1000 datasets (root-mean variance: RMVAR), the standard deviation of the bias across all 1000 datasets (Monte Carlo standard error: MCSE), the 95% confidence interval coverage (proportion of estimated confidence intervals that included the true value), and the type-I error (when the null hypothesis is true) or statistical power (when the null hypothesis is false). Our main results are for sample sizes of 500, but we evaluated alternative sample sizes (n=100, 2000, 5000) in sensitivity analyses. Sample R code for simulation analyses is given in the Supplementary Material.

RESULTS
*Large sample simulation*
In large samples (N=100,000) in which the assumptions of WQS hold, WQS regression and quantile g-computation give identical, unbiased estimates of the overall exposure effect and the weights, though WQS regression yields a smaller test statistic (larger p-value) of the null hypothesis test due to the use of sample splitting (**Table 2**). Weights are displayed graphically in the supplement (**Figures S1, S2**).

**Table 2: Single simulation demonstrating equivalence between WQS and quantile g-computation in large samples (N=100,000) when all exposures have effects in the same direction (true $\psi = 5.0$ true weights = 0.5, 0.25, 0.15, 0.1).**

|  | Mixture effect | | Estimated weights | | | |
|---|---|---|---|---|---|---|
| **Method** | $\hat{\psi}$ | t-statistic | $\hat{w_1}$ | $\hat{w_2}$ | $\hat{w_3}$ | $\hat{w_4}$ |
| **WQS**[a] | 5.00 | 806 | 0.50 | 0.25 | 0.15 | 0.10 |
| **Q-gcomp**[b] | 5.00 | 884 | 0.50 | 0.25 | 0.15 | 0.10 |

[a]WQS regression (R package "gWQS" defaults)
[b]Quantile g-computation (R package "qgcomp" defaults)

*Small and moderate sample simulations*
*Validity under the null hypothesis $\psi = 0$ when 1) exposures have no effect (Scenario 1) or 2) exposures have counteracting effects (Scenario 2)*: Under the null hypothesis when no exposures have an effect on the outcome, both quantile g-computation and WQS regression yield type 1 error rates close to the alpha level of 0.05 at all values of $d$ we examined (**Table 3**). When some exposures cause the outcome, but there is no overall exposure effect due to counteracting effects of exposures ($\beta_1 = -\beta_2 = 0.25$, quantile g-computation provided a valid test of the null, whereas WQS regression was biased away from the null, and this bias appeared to increase with the number of "noise" exposures included in the model, which led to type I error rates >90%. Results were similar at N=100 (**Table A1**). We note that the results when $\psi = 0$ do not depend on whether effects are constrained to be positive or negative in WQS regression, due to symmetry of the linear model.

**Table 3: Validity of WQS regression and quantile g-computation under the null (no exposures affect the outcome, or exposures counteract) and non-null estimates when directional homogeneity holds, 1,000 simulated samples of N=500. Corresponding estimates for N=100 are provided in Table A1.**

| Scenario | Method | d[a] | Truth[b] | Bias[c] | MCSE[d] | RMVAR[e] | Coverage[f] | Power/ Type 1 error[g] |
|---|---|---|---|---|---|---|---|---|
| 1. Validity under the null, no exposures are causal |  | 4 | 0 | 0.00 | 0.09 | 0.09 | 0.95 | 0.05 |
|  | WQS[h] | 9 | 0 | -0.01 | 0.13 | 0.12 | 0.94 | 0.06 |
|  |  | 14 | 0 | -0.01 | 0.15 | 0.15 | 0.95 | 0.05 |
|  | Q-gcomp[i] | 4 | 0 | 0.00 | 0.08 | 0.08 | 0.94 | 0.06 |

| Scenario | Method | Exposures[a] | True $\psi$[b] | Bias[c] | MCSE[d] | RMV[e] | Coverage[f] | Power/Type1[g] |
|---|---|---|---|---|---|---|---|---|
| | | 9 | 0 | 0.00 | 0.12 | 0.12 | 0.95 | 0.05 |
| | | 14 | 0 | -0.01 | 0.16 | 0.15 | 0.95 | 0.05 |
| 2. Validity under the null, causal exposures counteract | WQS[h] | 4 | 0 | 0.32 | 0.08 | 0.08 | 0.02 | 0.98 |
| | | 9 | 0 | 0.41 | 0.11 | 0.11 | 0.04 | 0.96 |
| | | 14 | 0 | 0.46 | 0.14 | 0.14 | 0.09 | 0.91 |
| | Q-gcomp[i] | 4 | 0 | 0.00 | 0.09 | 0.09 | 0.95 | 0.05 |
| | | 9 | 0 | 0.00 | 0.13 | 0.13 | 0.96 | 0.04 |
| | | 14 | 0 | -0.01 | 0.16 | 0.16 | 0.96 | 0.04 |
| 3. Validity under single non-null effect | WQS[h] | 4 | 0.25 | 0.07 | 0.07 | 0.07 | 0.83 | 1.00 |
| | | 9 | 0.25 | 0.15 | 0.10 | 0.10 | 0.67 | 0.98 |
| | | 14 | 0.25 | 0.21 | 0.14 | 0.13 | 0.57 | 0.94 |
| | Q-gcomp[i] | 4 | 0.25 | 0.00 | 0.08 | 0.08 | 0.94 | 0.88 |
| | | 9 | 0.25 | 0.00 | 0.12 | 0.12 | 0.95 | 0.52 |
| | | 14 | 0.25 | -0.01 | 0.16 | 0.15 | 0.95 | 0.36 |
| 4. Validity under all non-null effects with directional homogeneity | WQS[h] | 4 | 0.25 | -0.06 | 0.10 | 0.09 | 0.87 | 0.58 |
| | | 9 | 0.25 | -0.09 | 0.13 | 0.13 | 0.87 | 0.26 |
| | | 14 | 0.25 | -0.10 | 0.17 | 0.15 | 0.88 | 0.19 |
| | Q-gcomp[i] | 4 | 0.25 | 0.00 | 0.08 | 0.08 | 0.95 | 0.86 |
| | | 9 | 0.25 | 0.00 | 0.12 | 0.12 | 0.95 | 0.55 |
| | | 14 | 0.25 | -0.01 | 0.15 | 0.15 | 0.95 | 0.37 |

[a]Total number of exposures in the model
[b]True value of $\psi$, the net effect of the exposure mixture
[c]Estimate of $\psi$ minus the true value
[d]Standard deviation of the bias across 1000 iterations
[e]Square root of the mean of the variance estimates from the 1000 simulations, should equal MCSE if the variance estimator is unbiased
[f]Proportion of simulations in which the estimated 95% confidence interval contained the truth.
[g]Power when the effect is non-null (scenarios 3 and 4), otherwise (scenarios 1 and 2) is the type 1 error rate (false rejection of null), which should equal alpha (0.05 here) under a valid test.
[h]WQS regression (R package "gWQS" defaults)
[i]Quantile g-computation (R package "qgcomp" defaults)

*Validity when $\psi \neq 0$ (Scenarios 3, 4)*: When all exposure effects were either positive or null, quantile g-computation provided unbiased effect estimates for the overall exposure effect, whereas WQS regression was biased away from the null (**Table 3**). For a single causal exposure (scenario 3), weighed quantile sum regression was more powerful, with power > 90% for up to 14 total exposures, but 95% confidence intervals had poor coverage (57-83%), while quantile g-computation provided valid confidence intervals, but at reduced power. When all exposures had positive effects (scenario 4), quantile g-computation was more powerful, and WQS regression was biased towards the null. Results were generally similar at N=100, but root-mean variance was lower than the Monte Carlo standard error for WQS regression at N=100, indicating the standard error estimates were biased to be too small (**Table A1**).

*Validity under co-pollutant and unmeasured confounding (Scenarios 5, 6)*: Under negative co-pollutant confounding, quantile g-computation was unbiased for all examined values of the total exposure effect and correlation between the causal exposures (**Figure 1**). Results were similar for both N=100 and N=500, and across the number of "noise" exposures included (**Figures S3-S7**). WQS regression was biased at all studied levels of confounding, and the bias increased with the strength of the negative confounder-outcome association and decreased with the correlation between the two exposures. After noting that the standard error of quantile g-computation estimates of $\psi$ <u>appeared to decrease</u> at higher levels of exposure correlation, we repeated a simpler version of this scenario at higher correlations (up to 0.9, N=500, $d$=4, $\beta_1 = \psi = 0.25$) for quantile g-computation only and observed that, following intuition, the confidence interval width (3.92*standard error) of $\beta_1$ increased with exposure correlation, but, counterintuitively, the confidence interval width of the overall effect $\psi$ *decreased* with exposure correlation (**Figure 2**).

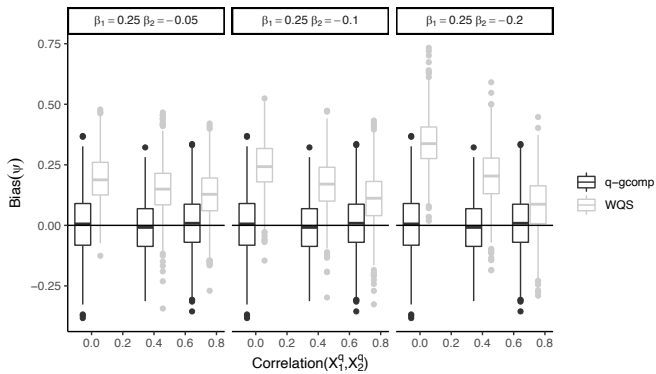

**Figure 1. Scenario 5: Impact of co-pollutant confounding on the bias of the overall exposure effect estimate (N=500, d=9) for quantile g-computation (q-gcomp) and WQS regression (WQS) at exposure correlations ($\rho_{X_1,X_2}$ of 0.0, 0.4, and 0.75) and varying total effect sizes ( $\psi = \beta_1 + \beta_2 \in 0.2, 0.15, 0.05$). Boxes represent the median (center line) and interquartile range (outer lines of box) and outliers (points outside of the 1.5*IQR length whiskers) across 1,000 simulations. Corresponding figures for d=4,14 and N=100 ($d$=4,9,14) are provided in Figures S3-S7.**

With increasing numbers of "noise" exposures, bias due to unmeasured confounding increased with WQS regression, whereas bias was stable with increasing "noise" exposures across all sample sizes studied for quantile g-computation (**Figure 3**). After noting that the magnitude of bias for WQS regression seemed to depend on sample size, we increased the sample size in our analysis (up to 5000). The difference between the two approaches diminished as sample size increased but was present at all sample sizes.

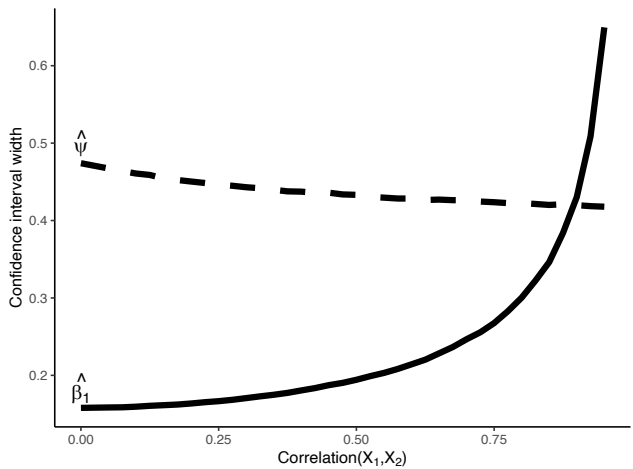

**Figure 2. Scenario 5: Impact of co-pollutant confounding on the confidence interval width of individual exposure estimates ($\beta$) and the overall exposure effect estimate ($\psi$) for quantile g-computation (q-gcomp) under exposure correlations ($\rho_{X_1,X_2}$) from 0.0 to 0.9 (N=500, $d$=9, $\psi = \beta_1 = 0.25$).**

*Validity under non-linearity and non-additivity (Scenarios 7 and 8)*: When exposure effects were non-linear and non-additive, WQS regression (even allowing for quadratic effects of the exposure index) yielded biased estimates of the quadratic exposure effects (the main effect of all exposures and the effect of all exposures, squared), whereas quantile g-computation yielded unbiased estimates of exposure effects and associated variances (**Table 4**). Quantile g-computation also provided more precise estimates.

Table 4: Validity of WQS regression and quantile g-computation under non-null estimates when directional homogeneity holds, individual exposure effects are non-additive, and the overall exposure effect includes terms for linear ($\psi_1$) and squared ($\psi_2$) exposure (e.g. quadratic polynomial), 1,000 simulated samples of N=500. Corresponding estimates for N=100 are provided in Table A2.

| Scenario | Method | $d^a$ | Bias[b] $\psi_1$ | $\psi_2$ | MCSE[c] $\psi_1$ | $\psi_2$ | RMVAR[d] $\psi_1$ | $\psi_2$ |
|---|---|---|---|---|---|---|---|---|
| 7. Validity when the true exposure effect is non-additive/non-linear | WQS[e] | 4 | 0.21 | -0.07 | 0.34 | 0.11 | 0.31 | 0.10 |
| | | 9 | 0.21 | -0.07 | 0.73 | 0.24 | 0.64 | 0.21 |
| | | 14 | 0.13 | -0.04 | 1.12 | 0.37 | 1.02 | 0.34 |
| | Q-gcomp[f] | 4 | -0.01 | 0.00 | 0.13 | 0.04 | 0.13 | 0.04 |
| | | 9 | 0.00 | 0.00 | 0.16 | 0.03 | 0.16 | 0.04 |
| | | 14 | 0.00 | 0.00 | 0.19 | 0.04 | 0.18 | 0.04 |
| 8. Validity when the overall exposure effect is non-linear due to underlying non-linear effects | WQS[e] | 4 | -0.20 | 0.07 | 0.31 | 0.12 | 0.31 | 0.10 |
| | | 9 | -0.15 | 0.07 | 0.61 | 0.22 | 0.61 | 0.20 |
| | | 14 | -0.11 | 0.05 | 0.97 | 0.34 | 0.98 | 0.32 |
| | Q-gcomp[f] | 4 | -0.01 | 0.00 | 0.15 | 0.04 | 0.15 | 0.04 |
| | | 9 | 0.00 | 0.00 | 0.18 | 0.05 | 0.18 | 0.05 |
| | | 14 | 0.00 | 0.00 | 0.20 | 0.05 | 0.20 | 0.05 |

[a]Total number of exposures in the model
[b]Estimate of $\psi_1$ or $\psi_2$ minus the true value
[c]Standard deviation of the bias across 1000 iterations
[d]Square root of the mean of the variance estimates from the 1000 simulations, should equal MCSE if the variance estimator is unbiased
[e]WQS regression (R package "gWQS" defaults, allowing for quadratic term for total exposure effect)
[f]Quantile g-computation (R package "qgcomp" defaults, including interaction term between $X_1$ and $X_2$ (scenario 7) or a term for $X_1 X_1$ (scenario 8) as well as quadratic term for total exposure effect)

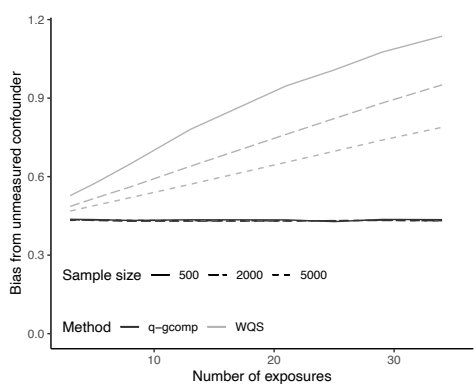

Figure 3. Scenario 6: Impact of unmeasured confounding on the bias of the overall exposure effect estimate (mean across 1,000 simulations, N=500, 2,000 or 5,000) for quantile g-computation (q-gcomp) and WQS regression (WQS) with confounder correlation ($\rho_{X,C}$) of 0.75, $\beta_C = 0.5$ and varying the total number of "noise" exposures ($d \in 4, 7, 9, 14, 22, 29, 35$). Note that all lines for quantile g-computation are overlapping and indicate unmeasured confounding bias is similar across sample sizes and number of exposures.

DISCUSSION

One of the remarkable and unique aspects of WQS regression in the context of exposure mixtures is that it specifically estimates a joint effect of the entire mixture: the effect of increasing all exposures by a single quantile. We show this approach to mixtures can leverage the existing correlation among exposures, where it may be difficult, if not impossible, to identify independent effects among correlated exposures (12). Further, for exposures such as hazardous air pollutants (33), phthalates and parabens (34, 35), and metals (36), feasible interventions to address any individual exposure would likely affect multiple exposures, leading to natural interest in joint effects. We built on this framework of quantized exposures and a single joint effect by combining these elements with existing causal inference approaches. While this bridge is enabled by the large sample equivalence between WQS regression and quantile g-computation when the necessary assumptions of WQS regression are met, we were primarily motived by developing an approach for estimating effects of mixtures in realistic sample sizes and when these crucial assumptions may not be met. To these ends, quantile g-computation maintains the simple inferential framework of WQS, while providing effect estimates that are robust to routine problems of exposure mixtures.

Under the directional homogeneity, linearity, and additivity assumptions, we demonstrate that WQS regression can be interpreted as an ordinary least-squares linear regression model with a coefficient that corresponds to the expected change in an outcome from a simultaneous increase in all exposures by a single quantile. Further, the approach is quite powerful when a single exposure has an effect in the expected direction. However, this power seems illusory since it results from effect estimates that are biased away from the null and, in small samples, downwardly biased variance estimates. Further, we showed that if the multiple exposures affect the outcome, then WQS regression is biased downward and has reduced power relative to quantile g-computation. For investigators choosing between WQS and quantile g-computation, we note that quantile g-computation was less biased than WQS under every scenario we examined using simulations. Any set of simulations cannot be considered exhaustive, as individual substantive contexts vary widely. For example, in some scenarios WQS demonstrated lower variance, suggesting that there may be a bias-variance tradeoff when choosing between the methods under certain contexts (e.g. when no exposures have an effect, WQS may be more precise). We emphasize, however, that the information driving the choice between methods is not often known in the context of exposure mixtures, where individual effects are rarely known for all mixtures components, a priori. In such settings, quantile g-computation is less prone to bias and appears robust in the sense that it does not produce or enhance spurious results when the assumptions of WQS are incorrect.

One aspect we did not address in our simulations was the use of sample splitting in WQS, where the data are divided into training/validation sets, and the weights are first estimated in the training set and then applied in the validation set to estimate the overall mixture effect. In small/moderately sized studies, sample splitting has been useful in other domains (such as in neural network-based machine learning) to avoid overfit, and in large samples sample splitting will not affect inference for consistent estimators of the weights under directional homogeneity (as demonstrated in our large sample simulation). The use of sample splitting allows the constraint on the weights to be on the other side of the null from the overall mixture effect. In the WQS R package (with default 40%/60% training/validation split), one can also estimate the weights and the overall mixture effect in the same data, although such a practice is generally discouraged. We repeated a limited version of our simulation scenario 1 in which we did this (only N=500), and found that estimating the weights in the same data as the exposure effect is estimated in leads to null hypothesis tests for WQS regression (but not quantile g-computation, which does not utilize sample splitting) that are no longer valid with type-I error rates >80% under some scenarios and bias that grows with the number of "noise" exposures included (**TABLE A3**).

Contrasted with quantile g-computation, WQS regression (that is, the implementation in the "gWQS" package) has three features: 1) a non-negativity/non-positivity constraint on weights (necessitating the directional homogeneity assumption), 2) bootstrap sampling to estimate weights, and 3) sample splitting which utilizes a training and validation set for calculation of the weights. As noted by Carrico et al, several of these elements have demonstrated some utility in algorithmic learning, suggesting that their use may improve some aspects of analysis over linear models (15). However, it appears that the non-negativity constraint (which implies the directional homogeneity assumption) is responsible for the bias away from the null we observed when directional homogeneity was violated or when there was a single causal exposure (Scenarios 2, 3, and 5). Furthermore, we speculate that bootstrap sampling is responsible for the bias towards the null when there were multiple causal exposures (Scenario 4). This latter result occurs because weights are 1 or 0 for every bootstrap sample simply due to sampling variability, so we posit there is some bias towards 0.5 for the weights, which would equate to a bias towards the null in this scenario. This scenario suggests that, even if we could *a priori* change the coding of exposures such that all effects are in the same direction, WQS regression would still yield biased results with confidence intervals that are too narrow. We also speculate that sample splitting is responsible for reduced power of WQS regression relative to quantile g-computation in some scenarios.

Exposure correlation in an epidemiologic context can indicate confounding, and we demonstrated that the non-negativity constraint can magnify confounding bias. This result is intuitive. Using the example of fish consumption and cognitive functioning: it is likely that the consumption of docosahexaenoic acid (DHA) from fish improves cognitive health, but fish are also sources of mercury exposure. If we analyzed these exposures in relation to cognitive functioning together under a non-positivity constraint, the effect of DHA would likely be forced to be near zero. This would be roughly equivalent to leaving DHA out of the model altogether, which would result in confounding in the negative direction for the effect of mercury. If we reversed the constraints, we would over-estimate the benefits of fish consumption because we would induce confounding in the opposite direction. Other examples of such phenomena are abundant in dietary epidemiology, where beneficial and harmful components frequently exist in the same food. Similarly, well water may include beneficial aspects such as trace essential minerals (37) and harmful constituents such as metals (38). More generally, however, we found little apparent benefit with respect to bias or variance at imposing the directional homogeneity assumption when estimating the effects of mixtures, even when it was true. By avoiding the non-negativity (or non-positivity) constraints, quantile g-computation can give a more realistic estimate of the effect of the mixture as a whole. Further, while our analysis focuses on the overall mixture effect, our approach also yields $\beta$ coefficients for an underlying generalized linear model, which can be interpreted as adjusted, independent effect sizes for quantized exposures (which will suffer variance inflation in adjusted analysis, like any other method that estimates independent effects). Further, the "mixture" under consideration could include only a subset of the measured exposures while controlling for other measured exposures. Such "conditional mixture effects" may be of interest in some settings where exposure sources can be identified that influence the levels of some, but not all components of the mixture.

Beyond our simulations in which the mixture was defined to include all simulated exposures, the question "how do we define a mixture?" is not clearly answered. However, our results provide some guidance. If we have correlated exposures that are all causal, then we usually ought to (if possible) include all correlated exposures in the model in order to avoid co-pollutant confounding (as in the fish consumption example and our Scenario 5 simulation). We should also account for interactions and non-linearities, if they exist (as in our Scenarios 7 and 8 simulations). Outside of co-pollutant confounding, however, we demonstrated that quantile g-computation is robust to the inclusion of non-causal "noise" variables, while bias due to unmeasured confounding actually increases with the inclusion of more "noise" variables for WQS regression. This phenomenon of increasing bias by including more covariates was also observed when no confounding existed and an increasing number of causal exposures were included, though the bias was towards the null in that setting (Scenario 4). Thus, the bias magnification observed in scenario 5 is not akin to "bias amplification" which is a phenomenon of predictably increased unmeasured confounding bias that occurs when examining independent effects of one exposure and adjusting for a strong correlate of that exposure (e.g. (39)). Of the two methods, quantile g-computation appears robust to varying definitions of "the mixture" and so seems appealing in the context of undefined mixtures where we may not have good prior knowledge of the constituent effects.

Notably, our results suggest that both quantile g-computation and WQS regression, through their focus on the effect of the mixture as a whole, do not necessarily suffer greatly from exposure correlation. Intuitively, this is true because, as exposures become more correlated, we gain more information on the expected effects of increasing every exposure simultaneously, in contrast with focusing on the effects of a single exposure while holding others constant.

Other methods could be used to estimate effects of the mixture as a whole. For example, if we fit quantile-g computation without using quantized exposures, the approach would yield an effect of increasing every exposure by one unit – this is useful when "one unit" is meaningful for all exposures. In that case, our approach would simply be "g-computation" (A.K.A the g-formula) (40), which has been previously used to estimate the impacts of hypothetical interventions in (among others) environmental and occupational (18, 20, 22) settings. More generally, g-computation provides a useful framework to estimate joint effects of an exposure mixture (4), especially when exposures vary over time (11) or when issues such as exposure measurement error may be important (40). Thus, our approach is potentially extensible to such scenarios and its utility is worthy of future study. There are examples of using such a framework with Bayesian Kernel Machine regression (BKMR) in a mixtures specific setting (41). Quantile g-computation is a simpler version of these approaches, but with the added strengths of being simple to implement and computationally frugal.

We posit that quantile g-computation (and the accompanying R package "qgcomp") provide a simple framework that allows a flexible approach to analysis of mixtures data when the overall exposure effect is of interest. Few methods explicitly estimate such effects. One alternative method is BKMR, which uses Gaussian process regression to fit a flexible function of the joint exposure set. In contrast, quantile g-computation forces the user to specify a parametric non-linear model, rather than assuming non-linearity by default. Whereas quantile g-computation allows estimation of parsimonious mixture-dose-response parameters, BKMR (and many other data-adaptive or machine learning approaches) does not yield a mixture-dose-response parameter that could be compared with quantile g-computation in terms of the bias/variance tradeoffs quantified in our analysis. Instead, BKMR (as implemented in the R package "bkmr") outputs a flexible prediction of the outcome at quantiles of all exposures, which cannot be expressed via a simple parametric function (which is required in order to assess bias of a mixture-dose-response parameter). Another way to understand this distinction is that quantile g-computation estimates the parameters of a joint marginal structural model, which quantify the average (or baseline confounder/modifier conditional) effects of modifying all exposures simultaneously. Interpretation of marginal structural models is well described in the epidemiologic literature (e.g. (42)). A limitation of this approach as implemented in the qgcomp R package is that if the underlying model is not smooth (e.g. if the dose response of an individual exposure is a step function as when using indicator functions of individual exposures), the marginal structural model (which in the current version of the package can only be a polynomial function of exposures) may not adequately capture the dose-response function. In such settings, standard g-computation may be employed, at a much higher computational and programming burden to the analyst. However, quantile g-computation may still provide useful approximations of overall effects of the mixture. Our simulations demonstrate that when the non-linearities are known, quantile g-computation is unbiased. Whether this holds in realistic settings when non-linearities are unknown will depend on the subject matter at hand and cannot be effectively explored via simulations without being tied more strongly to specific exposures and outcomes. We note that this is a potential limitation of parametric modeling in general and is not particular to our approach. However, when studying causal effects in general and mixture effects specifically, model specification must be accurate for accurate inference, which underscores the importance of allowing non-linear and non-additive effects of individual exposures (11). Understanding non-linearity in the overall exposure effect is crucial for understanding whether effects are mainly limited to a certain range of joint exposures within the mixture, which informs whether interventions might be most impactful if they focused solely on individuals in a specific exposure range.

The reliance of the qgcomp package on existing generalized linear model frameworks allows one to easily include a variety of flexible non-linear/non-additive features such as polynomial functions, splines, interaction terms or indicator functions on the individual quantized exposures (which are each described in the package documentation). While the utility of quantized exposures will depend on the context, the impact of characterizing rich exposure data as ordinal variables that are used as continuous regressors is broadly unknown. Our simulation did not assess the benefits and tradeoffs of using higher numbers of quantiles because the exposures were simulated on the quantized, rather than continuous basis and little could be learned by increasing the numbers of exposure levels in our simulation. We hypothesize that quantization confers some benefits of similar non-parametric approaches such as rank regression, which are robust to outliers, but may suffer from difficulty in extrapolating results to other populations and reduced power when a linear model fits the data well. This is an area worth more research in the exposure mixtures context, where skewed exposure distributions may result in unduly influential, extreme exposure values and non-linear effects may be common.

We propose a method that builds on the desirable, simple output from WQS regression, but is appropriate to use when the effects of exposure may be beneficial, harmful, or harmless. In scenarios where we may not be able to rule out confounding or we may be uncertain about the effect direction of some exposures in the mixture, quantile g-computation is a simple and computationally efficient approach to estimating associations between a mixture of exposures and a health outcome of interest. Thus, our approach may serve as a valuable tool for identifying mixtures with harmful constituents or informing interventions that may prevent or reduce multiple exposures within a mixture.

# APPENDIX

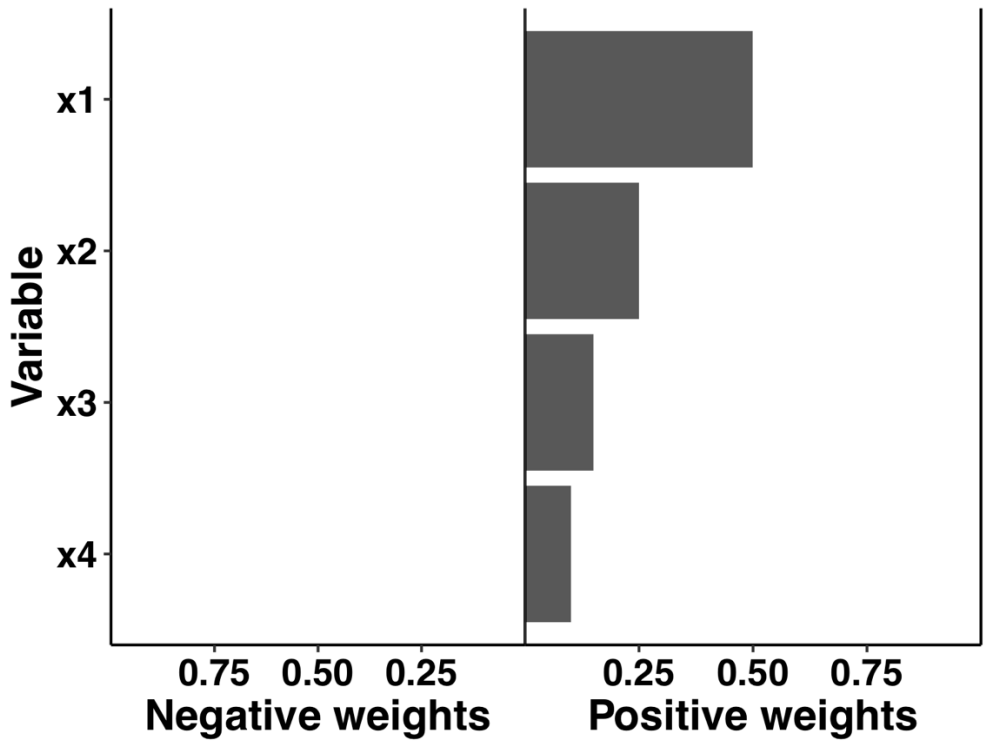

**Figure A1:** Large sample simulation results demonstrating the default graphical output of the "qgcomp" R package showing the point estimates of the exposure weights in quantile g-computation (when weights are estimable in a linear/additive model). Under a linear model when effects of exposures are all in the same direction, the weights correspond to the proportion of the total mixture effect that is due to each exposure.

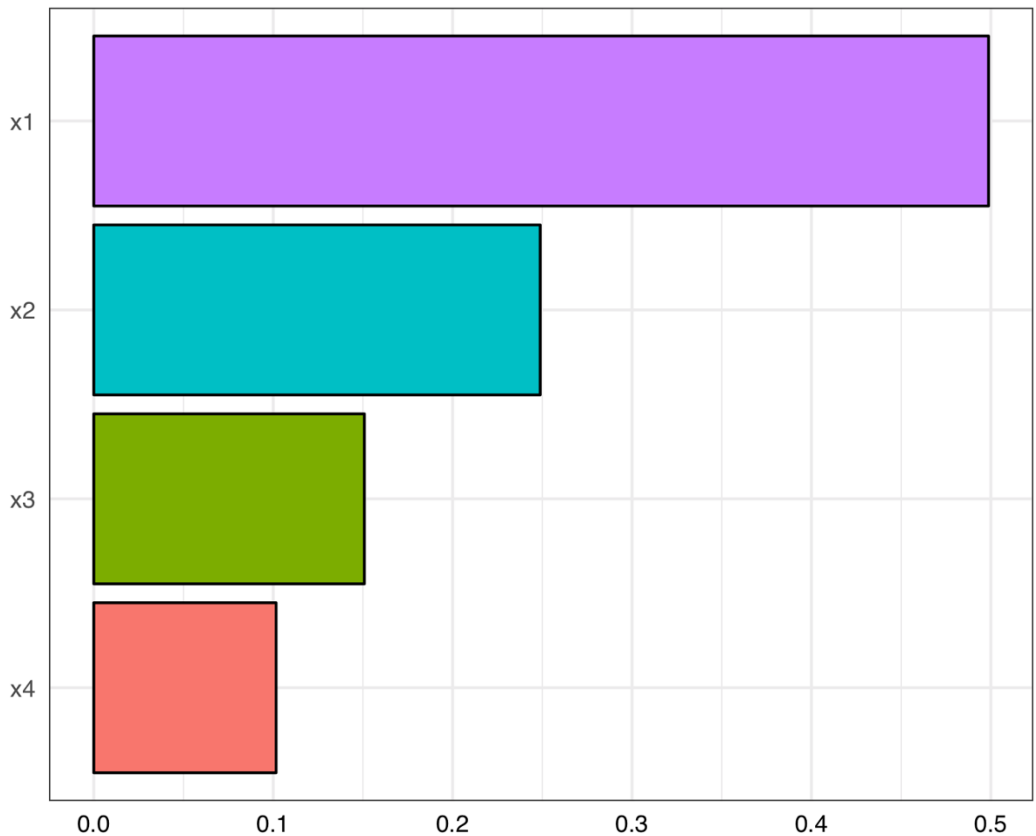

**Figure A2: Large sample simulation results demonstrating the default graphical output of the "gWQS" R package showing the point estimates of the exposures weights in WQS regression. Under a linear model when effects of exposures are all in the same direction, the weights correspond to the proportion of the total mixture effect that is due to each exposure.**

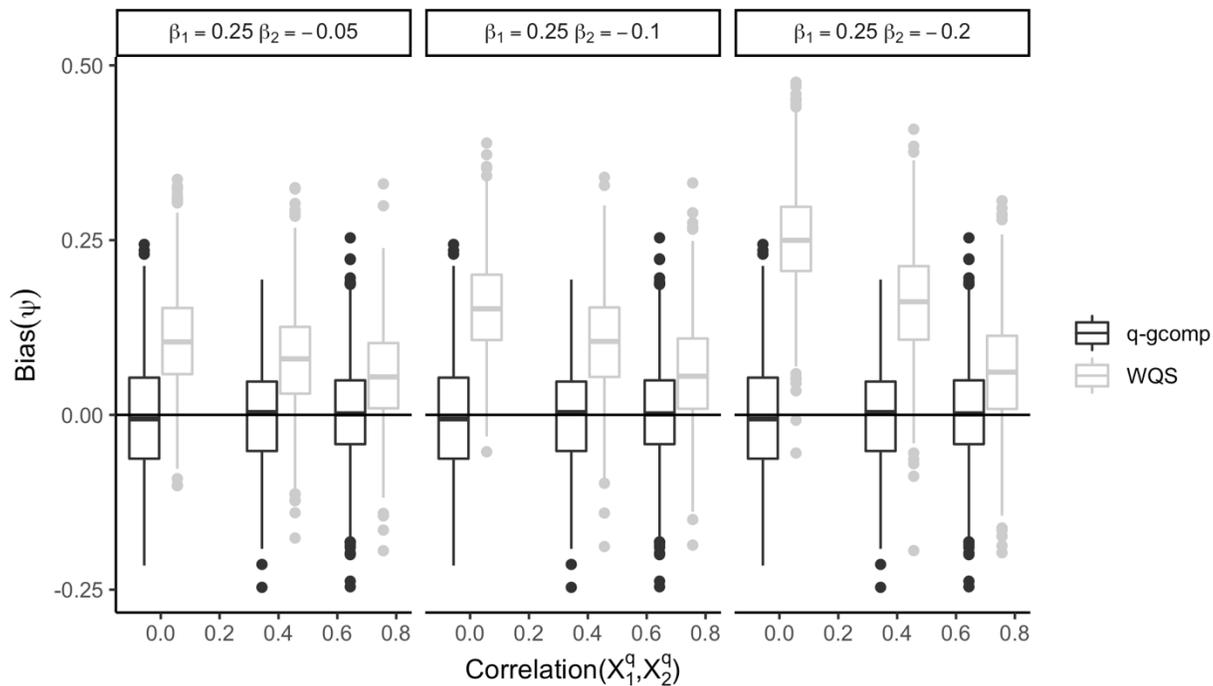

**Figure A3. Scenario 5: Impact of co-pollutant confounding on the bias of the overall exposure effect estimate (N=500, d=4) for quantile g-computation (q-gcomp) and WQS regression (WQS) at exposure correlations**

($\rho_{X_1X_2}$ of 0.0, 0.4, and 0.75) and varying total effect sizes ($\psi = \beta_1 + \beta_2 \in 0.2, 0.15, 0.05$). Boxes represent the median (center line) and interquartile range (outer lines of box) and outliers (points outside of the 1.5*IQR length whiskers) across 1,000 simulations.

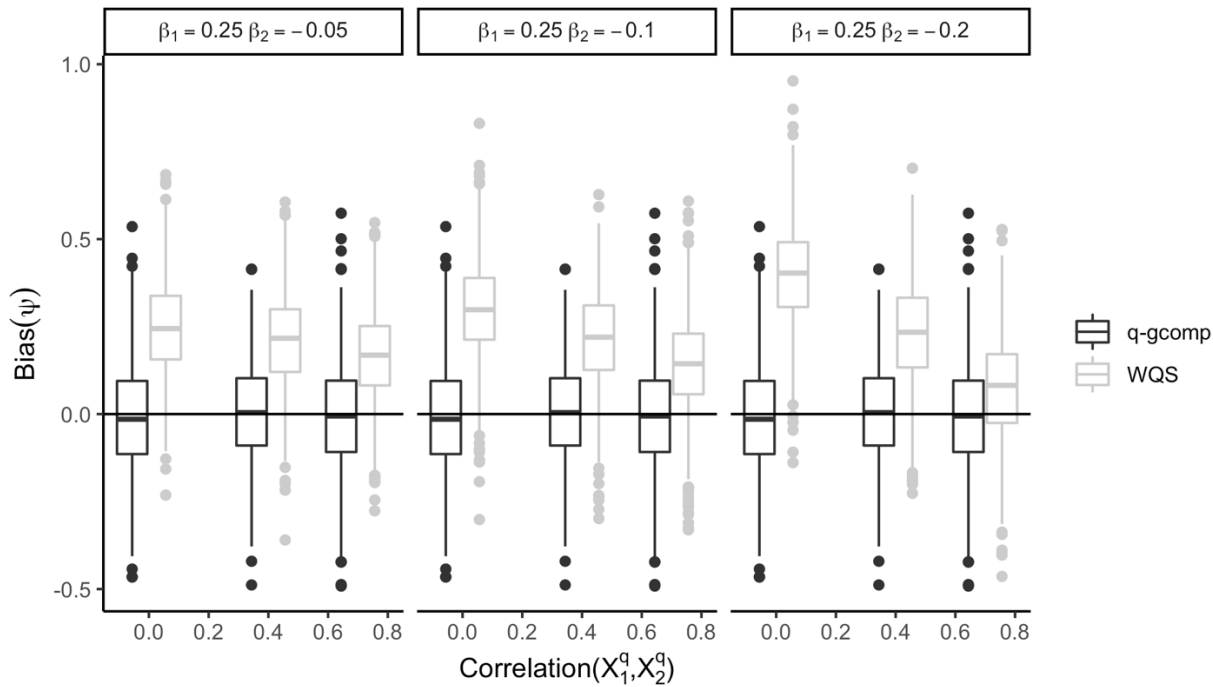

**Figure A4. Scenario 5: Impact of co-pollutant confounding on the bias of the overall exposure effect estimate (N=500, d=14) for quantile g-computation (q-gcomp) and WQS regression (WQS) at exposure correlations ($\rho_{X_1X_2}$ of 0.0, 0.4, and 0.75) and varying total effect sizes ($\psi = \beta_1 + \beta_2 \in 0.2, 0.15, 0.05$). Boxes represent the median (center line) and interquartile range (outer lines of box) and outliers (points outside of the 1.5*IQR length whiskers) across 1,000 simulations.**

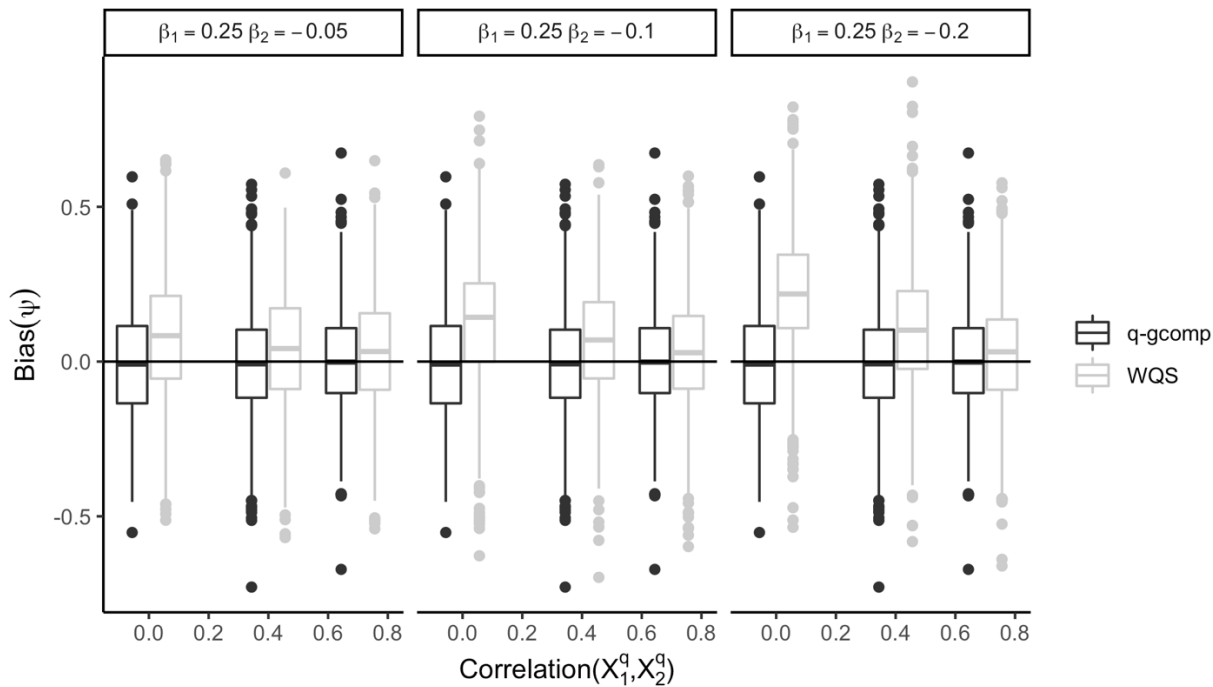

**Figure A5.**
**Scenario 5: Impact of co-pollutant confounding on the bias of the overall exposure effect estimate (N=100, d=4) for quantile g-computation (q-gcomp) and WQS regression (WQS) at exposure correlations ($\rho_{X_1X_2}$ of 0.0, 0.4, and 0.75) and varying total effect sizes ($\psi = \beta_1 + \beta_2 \in 0.2, 0.15, 0.05$). Boxes represent the median (center line) and interquartile range (outer lines of box) and outliers (points outside of the 1.5*IQR length whiskers) across 1,000 simulations.**

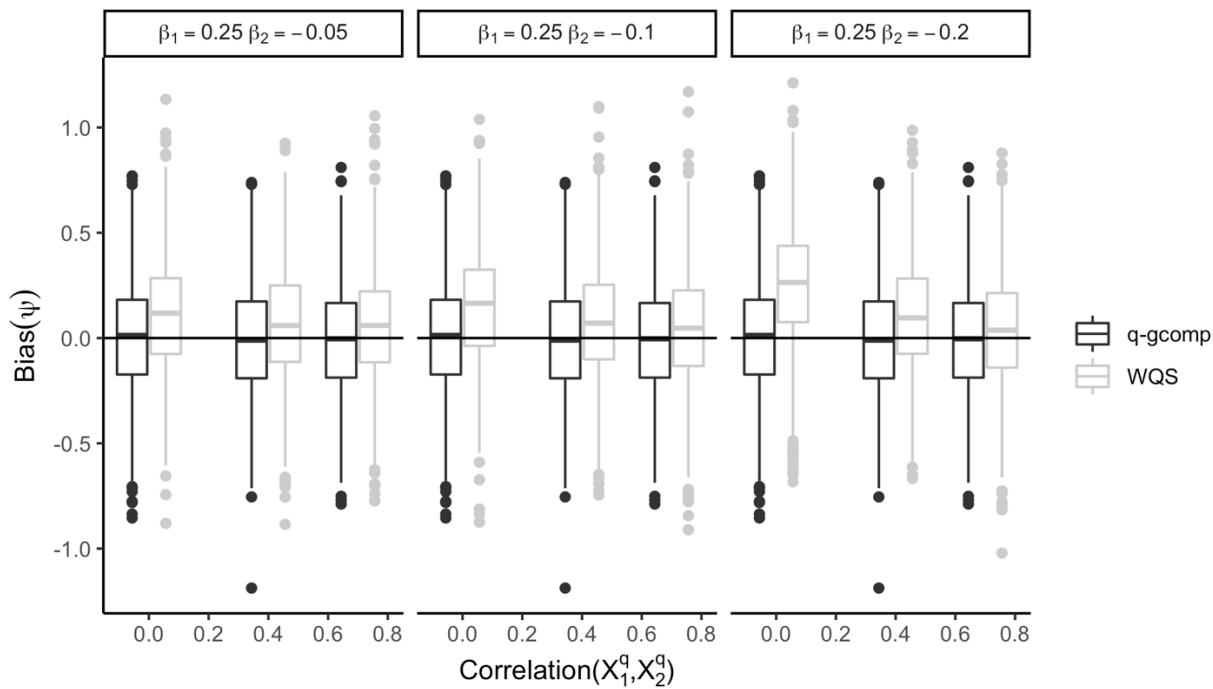

**Figure A6.**
**Scenario 5: Impact of co-pollutant confounding on the bias of the overall exposure effect estimate (N=100, d=9) for quantile g-computation (q-gcomp) and WQS regression (WQS) at exposure correlations ($\rho_{X_1X_2}$ of 0.0, 0.4, and 0.75) and varying total effect sizes ($\psi = \beta_1 + \beta_2 \in 0.2, 0.15, 0.05$). Boxes represent the median (center line) and interquartile range (outer lines of box) and outliers (points outside of the 1.5*IQR length whiskers) across 1,000 simulations.**

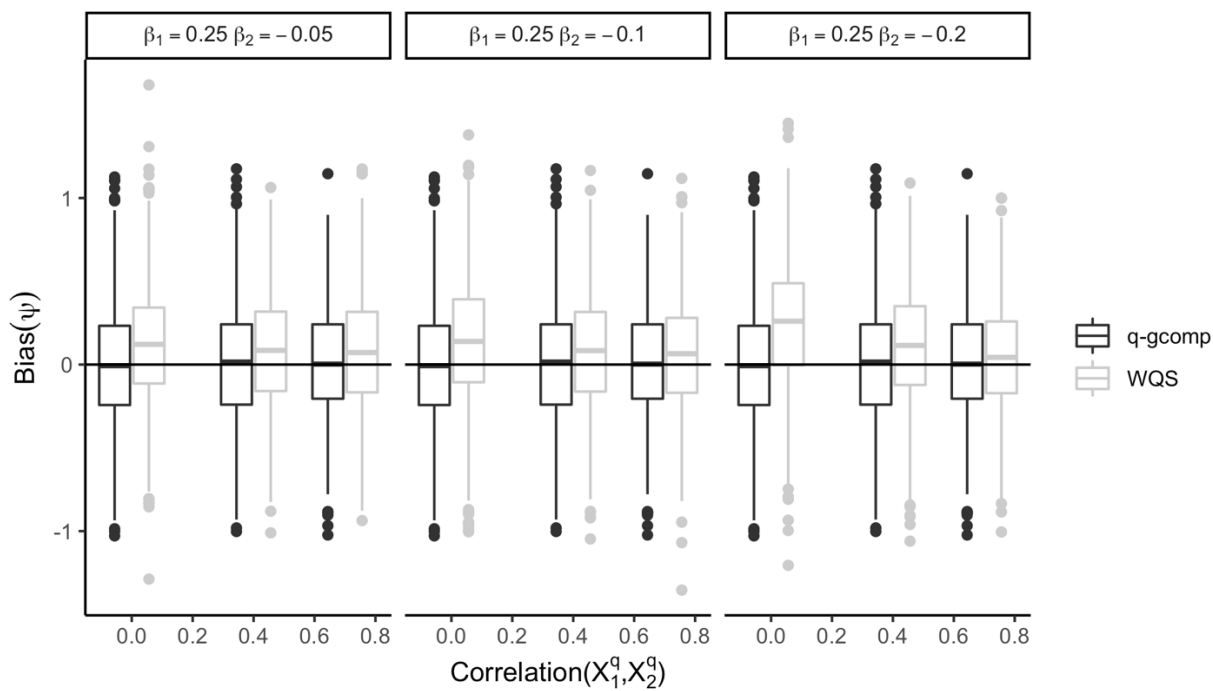

**Figure A7. Scenario 5: Impact of co-pollutant confounding on the bias of the overall exposure effect estimate (N=100, d=14) for quantile g-computation (q-gcomp) and WQS regression (WQS) at exposure correlations ($\rho_{X_1 X_2}$ of 0.0, 0.4, and 0.75) and varying total effect sizes ($\psi = \beta_1 + \beta_2 \in 0.2, 0.15, 0.05$). Boxes represent the median (center line) and interquartile range (outer lines of box) and outliers (points outside of the 1.5*IQR length whiskers) across 1,000 simulations.**

# APPENDIX TABLES

**Table A1: Validity of WQS regression and quantile g-computation under the null (no exposures affect the outcome, or exposures counteract) and non-null estimates when directional homogeneity holds, 1,000 simulated samples of N=100.**

| Scenario | Method | d[a] | Truth[b] | Bias[c] | MCSE[d] | RMVAR[e] | Coverage[f] | Power/ Type 1 error[g] |
|---|---|---|---|---|---|---|---|---|
| 1. Validity under the null, no exposures are causal | | 4 | 0 | 0.00 | 0.19 | 0.20 | 0.95 | 0.05 |
| | WQS[h] | 9 | 0 | 0.02 | 0.27 | 0.27 | 0.94 | 0.06 |
| | | 14 | 0 | 0.00 | 0.34 | 0.34 | 0.95 | 0.05 |
| | | 4 | 0 | 0.00 | 0.18 | 0.18 | 0.95 | 0.05 |
| | Q-gcomp[i] | 9 | 0 | 0.02 | 0.28 | 0.28 | 0.94 | 0.06 |
| | | 14 | 0 | 0.00 | 0.36 | 0.36 | 0.95 | 0.05 |
| 2. Validity under the null, causal exposures counteract | | 4 | 0 | 0.28 | 0.22 | 0.20 | 0.67 | 0.33 |
| | WQS[h] | 9 | 0 | 0.30 | 0.32 | 0.29 | 0.78 | 0.22 |
| | | 14 | 0 | 0.28 | 0.39 | 0.35 | 0.83 | 0.17 |
| | | 4 | 0 | 0.00 | 0.20 | 0.20 | 0.96 | 0.04 |
| | Q-gcomp[i] | 9 | 0 | 0.01 | 0.30 | 0.30 | 0.95 | 0.05 |
| | | 14 | 0 | -0.02 | 0.38 | 0.38 | 0.94 | 0.06 |
| 3. Validity under single non-null effect | | 4 | 0.25 | 0.04 | 0.21 | 0.19 | 0.92 | 0.37 |
| | WQS[h] | 9 | 0.25 | 0.09 | 0.29 | 0.27 | 0.91 | 0.27 |
| | | 14 | 0.25 | 0.07 | 0.36 | 0.34 | 0.92 | 0.19 |
| | | 4 | 0.25 | 0.00 | 0.18 | 0.18 | 0.95 | 0.29 |
| | Q-gcomp[i] | 9 | 0.25 | 0.02 | 0.28 | 0.28 | 0.94 | 0.16 |
| | | 14 | 0.25 | 0.00 | 0.36 | 0.36 | 0.95 | 0.12 |
| 4. Validity under all non-null effects with directional homogeneity | | 4 | 0.25 | -0.07 | 0.20 | 0.20 | 0.93 | 0.15 |
| | WQS[h] | 9 | 0.25 | -0.07 | 0.28 | 0.28 | 0.94 | 0.11 |
| | | 14 | 0.25 | -0.12 | 0.34 | 0.34 | 0.93 | 0.08 |
| | | 4 | 0.25 | -0.01 | 0.18 | 0.18 | 0.95 | 0.26 |
| | Q-gcomp[i] | 9 | 0.25 | 0.00 | 0.27 | 0.28 | 0.95 | 0.15 |
| | | 14 | 0.25 | 0.00 | 0.36 | 0.36 | 0.94 | 0.11 |

[a]Total number of exposures in the model
[b]True value of $\psi$, the net effect of the exposure mixture
[c]Estimate of $\psi$ minus the true value
[d]Standard deviation of the bias across 1000 iterations
[e]Square root of the mean of the variance estimates from the 1000 simulations, should equal MCSE if the variance estimator is unbiased
[f]Proportion of simulations in which the estimated 95% confidence interval contained the truth.
[g]Power when the effect is non-null, otherwise is the type 1 error rate (false rejection of null), which should equal alpha (0.05 here) under a valid test
[h]WQS regression (R package "gWQS" defaults)
[i]Quantile g-computation (R package "qgcomp" defaults)

Table A2: Validity of WQS regression and quantile g-computation under non-null estimates when directional homogeneity holds, individual exposure effects are non-additive, and the overall exposure effect includes terms for linear ($\psi_1$) and squared ($\psi_2$) exposure (e.g. quadratic polynomial), 1,000 simulated samples of N=100.

| Scenario | Method | d[a] | Bias[b] $\psi_1$ | $\psi_2$ | MCSE[c] $\psi_1$ | $\psi_2$ | RMVAR[d] $\psi_1$ | $\psi_2$ |
|---|---|---|---|---|---|---|---|---|
| 7. Validity when the true exposure effect is non-additive/non-linear | WQS[e] | 4 | 0.13 | -0.04 | 0.78 | 0.25 | 0.75 | 0.24 |
| | | 9 | -0.08 | 0.02 | 1.69 | 0.56 | 1.67 | 0.55 |
| | | 14 | -0.24 | 0.08 | 2.61 | 0.87 | 2.69 | 0.89 |
| | Q-gcomp[f] | 4 | 0.00 | 0.00 | 0.30 | 0.08 | 0.29 | 0.08 |
| | | 9 | 0.01 | 0.00 | 0.36 | 0.08 | 0.37 | 0.09 |
| | | 14 | -0.02 | 0.00 | 0.43 | 0.09 | 0.45 | 0.09 |
| 8. Validity when the overall exposure effect is non-linear due to underlying non-linear effects | WQS[e] | 4 | -0.21 | 0.07 | 0.74 | 0.24 | 0.77 | 0.25 |
| | | 9 | -0.25 | 0.09 | 1.68 | 0.55 | 1.71 | 0.56 |
| | | 14 | -0.39 | 0.13 | 2.63 | 0.87 | 2.77 | 0.91 |
| | Q-gcomp[f] | 4 | 0.00 | 0.00 | 0.34 | 0.10 | 0.34 | 0.10 |
| | | 9 | 0.00 | 0.00 | 0.41 | 0.10 | 0.41 | 0.10 |
| | | 14 | -0.01 | 0.00 | 0.46 | 0.11 | 0.49 | 0.11 |

[a]Total number of exposures in the model
[b]Estimate of $\psi_1$ or $\psi_2$ minus the true value
[c]Standard deviation of the bias across 1000 iterations
[d]Square root of the mean of the variance estimates from the 1000 simulations, should equal MCSE if the variance estimator is unbiased
[e]WQS regression (R package "gWQS" defaults, allowing for quadratic term for total exposure effect)
[f]Quantile g-computation (R package "qgcomp" defaults, including interaction term between $X_1$ and $X_2$ (scenario 7) or a term for $X_1X_1$ (scenario 8) as well as quadratic term for total exposure effect)

**Table A3: Validity of WQS regression without sample splitting and quantile g-computation under the null (no exposures affect the outcome, 1,000 simulated samples of N=500).**

| Scenario | Method | df[a] | Truth[b] | Bias[c] | MCSE[d] | RMVAR[e] | Type 1 error[f] |
|---|---|---|---|---|---|---|---|
| 1. Validity under the null, no exposures are causal | WQS[g] | 4 | 0 | 0.06 | 0.06 | 0.07 | 0.86 |
| | | 9 | 0 | 0.15 | 0.09 | 0.09 | 0.65 |
| | | 14 | 0 | 0.25 | 0.10 | 0.11 | 0.38 |
| | Q-gcomp[h] | 4 | 0 | 0.00 | 0.08 | 0.08 | 0.06 |
| | | 9 | 0 | 0.00 | 0.12 | 0.12 | 0.05 |
| | | 14 | 0 | -0.01 | 0.16 | 0.15 | 0.05 |

[a]Total number of exposures in the model
[b]True value of $\psi$, the net effect of the exposure mixture
[c]Estimate of $\psi$ minus the true value
[d]Standard deviation of the bias across 1000 iterations
[e]Square root of the mean of the variance estimates from the 1000 simulations, should equal MCSE if the variance estimator is unbiased
[f]Type 1 error rate (false rejection of null), which should equal alpha (0.05 here) under a valid test
[g]WQS regression (R package "gWQS", validation parameter set to 0)
[h]Quantile g-computation (R package "qgcomp" defaults). Results are repeated from Table 3 in the main text for reference.